# Assessment of POS Owners' Awareness of Cybersecurity and Insider Threats in POS Kiosks Related Financial Crimes

A Project Report

By

Rawlings Fayeofori Fiberesima

A00020331

School of Information, Technology & Computing

American University of Nigeria

Supervisor: Dr. Samuel C. Utulu

May, 2022

# Assessment of POS Owners' Awareness of Cybersecurity and Insider Threats in POS Kiosks Related Financial Crimes

**Rawlings Fayeofori Fiberesima**

**BSc Report**

**Being a Senior Project Report submitted in**

**Partial fulfilment of the requirements for the award of**

**Degree of Bachelor in Information Systems.**

**Supervisor: Dr. Samuel Utulu**

**School of Information, Technology & Computing**

**American University of Nigeria**

**May, 2022.**

# DECLARATION

I Rawlings Fiberesima declare that this project "Assessment of POS Owners' Awareness of Cybersecurity and Insider Threats in POS Kiosks Related Financial Crimes" is my original work and has not been submitted to any other institution. Also, I declare that this report was written in my own words, except for cited works, whose credit was provided. This project was thoroughly inspected and carried out under the supervision of Dr. Samuel Utulu.

……………………………                                              …………………..

Student: Rawlings Fiberesima                                         Date

…….…………………….                                              …..………………

Supervisor: Dr. Samuel C. Utulu                                      Date

………………………………......                                       …………………………..

Program Chair: Professor Rao Narasimha Vajjhala                      Date

............................................                        ................................

Dean: Professor Abel Ajibesin                                        Date

# CERTIFICATION

This is to certify that Rawlings Fiberesima (A00020331), an undergraduate in Information Systems in the department of the School of Information Technology and Computing (SITC) has satisfactorily rounded up the research project required for the Bachelor of Science in Information Systems. The work contained in this research is genuine and has not been turned in for submission in any partial or full diploma or degree in this university or any other.

______________________                    ______________________

Rawlings Fiberesima                                       Dr. Samuel C. Utulu

Student                                                              Supervisor

## DEDICATION

I dedicate this project to God almighty and to the Fiberesima family for the continuous love and support.

## ACKNOWLEDGEMENT

The possibility of this project was as a result of the help and guidance of my supervisor Dr. Samuel Utulu. I am grateful for his ability to be patient with me and also ensure the research carried out was of required standard. I also acknowledge the stakeholders involved for providing me with insight required for this project.

# ABSTRACT

The introduction of point of sales (POS) technologies as a payment system was welcoming. It constitutes one of the major breakthroughs in the efforts made to rejuvenate the global financial systems. However, like other information technology (IT) based financial systems, the POS also poses some cybersecurity security threats. The unique thing about the POS is that the main cybersecurity threats it poses to users are not IT based. It follows that the threats are not usually connected to hacking or unpermitted access. The threats are rather connected to the mode of operation of POS kiosks, particularly as experienced in most parts of Nigeria. The mode of operation exposes users' cards to possible cloning. It could also enable those operating POS kiosks to copy sensitive automatic teller machine (ATM) cards numbers and CCV numbers which could give them unauthorized access to users' accounts. While this are possible realities, the questions that come to mind is: are POS users aware of this cybersecurity threats and how do they prevent the threats from playing out? Therefore, the study was carried out to find out POS users' awareness of the cybersecurity threats including possible insider threats that are connected to the use of POS in kiosks for providing financial services. The study adopted the case study research method and drew its conclusions on the questions raised from data derived from interviews that were conducted with those that use POS in kiosks in the Yola metropolis. The findings of the result show that POS kiosks are prone to insider related cybercrimes. Although owners of POS kiosks make efforts to ensure that those that they employ to run their POS kiosks are credible, their efforts are still minimal and could be breached. We recommend that POS kiosks owners should pay more attention to issues related to insider related cybercrime that could be perpetrated by POS kiosks personnel.

## Table of Contents

# CHAPTER ONE
# INTRODUCTION

## 1.1 Background to the Study

There are several instances of financial crimes committed by bank employees. Similar incidents about financial crimes that bank employees helped perpetrate exist. (Hasham et al 2019). The evidence suggests that persons who are trusted with positions that demand honesty, integrity, and dependability occasionally break public trust. This has led to an increase in the quantity of research regarding insiders' involvement in bank-related crimes, their function in the commission of financial crimes, and most recently, cybercrimes. (Cummings et al 2012). Studies on insider crimes in the banking industry go back to the 1990s, a time when institutions had not yet implemented and utilized IT. (Randazzo et al 2005; Stevens. 1989). The majority of insider-related financial crimes were committed using stolen identities, according to a key finding from the early research that were published in the literature. This is due to the fact that the insiders involved had privileged access to individuals' personal information and were able to commit insider-related financial crimes including the provision of privileged information about bank users to offenders. Such details typically provide the offender a decisive advantage over bank customers.

In Nigeria, the problem of insider in the perpetration of financial crimes predates the adoption of IT by commercial banks and other financial institutions. It was not until the 2000s that inside related financial crimes became one of the many cybersecurity threats that commercial banks and other financial institutions had to deal with. This is following the adoption of IT and implementation of the electronic banking systems in the country (Ezeoha, 2005). According to Adegboye (2009) insider related financial crimes has resulted to situations in which the Central Bank of Nigeria revoked the license of banks in the country. Ezeoha (2005) demands for adequate regulation to control banking activities so as to reduce or eliminate the occurrence of insider-aided financial crimes. This is so because victims frequently struggle to recover their lost finances. In Nigeria, the problems associated with insider-assisted financial crimes have not diminished. Recent papers still show that the occurrence is common and becoming more complex. (Tade, 2021; Samuel et al 2018).

A study by Shaikh & Karjaluoto (2015) demonstrates that the usage of mobile devices has spread internationally. According to their data, there are more locations where mobile devices are used to deliver banking services and products than there are locations where they are not. Both banks and consumers may benefit greatly from using mobile devices to deliver banking goods and services. However, research on the factors that influence the adoption of the various mobile financial service systems have been documented in the literature. Numerous studies revealed that trust and security were key drivers of the adoption of mobile device-based financial goods and services, along with consumer happiness, internet accessibility, and the caliber of financial services. (Kasim et al. 2022; Geebren, et al. 2021; Sharma & Sharma, 2019; Zhang et al. 2018). It appears that cybersecurity concerns have an impact on the adoption of mobile devices to deliver financial goods and services.

Consequently, there are revelations in the literature that POS are useful to sustaining small businesses that is those that use it as payment systems and that POS helps to create job opportunities in developing countries (Prihatiningtias & Wardhani, 2021; Ndung'u, 2018). The adoption of POS, particularly in Nigeria, also faces some financial crime related challenges (Oyemakara, 2020; Williams et al. 2018). Many of the POS use financial crimes have been reported in the literature except those that are connected to insider related cybersecurity crimes. The implication of this study is devotion to assessing a new form of insider related cybersecurity crimes that are perpetrated due to the increase in the use of mobile devices for financial services.

## 1.2 Statement of the Problem

The use of POS has become very popular across Nigeria. POS is used both in urban and rural centers and has become a very reliable way for receiving and sending money. It is a category of mobile device based financial service delivery system operated by banks. The modes of operations of POS however, requires that people who may not have training in financial service management operate them. This is evident in departmental stores and small and medium enterprises that use POS as payment systems. It is also evident among new financial businesses where operators use POS to provide financial services that enable users withdraw or deposit money from and into their bank accounts. Given the increase in the use and effectiveness of POS in financial service management, many studies have been carried out with regards to the adoption and use of POS to provide financial services. The studies however, are mainly based on factors that promote and hinder the adoption and use of POS for financial services (e.g. Williams et al. 2018). Another theme that is popular in the literature on the use of POS for financial services is on the challenges of adopting and using POS (e.g. Onyemakara, 2020).

In this category of scholarly published works, authors look at the challenges related to frequency of incomplete transactions and the difficulty of rectifying problems connected to the use of POS for financial services by its owners in banks. There are also studies that are based on assessing cybersecurity issues that are connected to the use of POS for financial services. The main theme treated by studies in this category is usually based on looking into how culprits have unauthorized access into accounts that the POS are connected to due to the carelessness of those managing POS financial services. There are also studies in Nigeria that look into how armed robbers attack POS kiosks and go away with the stacks of cash that they have within the kiosks. The first problem with the themes that are treated by scholars and reported in the literature is that they represent the knowledge that are available to stakeholders that are involved in the use of POS for financial services. The implication is that there are many more issues that are connected to the use of POS for financial services that stakeholders do not have knowledge about. The primary problem which was treated in this study is that which is connected to cybersecurity related threats that are perpetrated by insiders in POS kiosks. This is considered pertinent because of the increase in the number of POS kiosks that render financial services, increase in the number of POS kiosks users and the dynamic nature of financial services related cybersecurity crimes. The study focuses on the awareness of POS kiosks owners of the cybersecurity related financial crimes that may perpetrated by insiders in POS kiosks. The study focuses on this problem because it is critical and can lead to other forms of cybersecurity threats that may affect the bank accounts and savings of potential victims.

## 1.3 Aims of the Study

Consequently, the broad objective set for this study is to assess the level of awareness among POS kiosks owners about the possibility that insider threats in POS kiosks can lead to cybersecurity issues. In order to be able to reach the broad objectives, three specific objectives were raised. The specific objectives are to:

- Find out the level of awareness among owners of POS kiosk owners that ensures that they serve as guardians against insider related financial threat in POS kiosks.
- Ascertain if POS Kiosks owners and customers are aware that POS attendants are likely offenders who may perpetrate insider related financial crimes in POS kiosks?
- Determine insider related financial crimes that have been experienced by POS kiosks owners and how are they connected to the business processes adopted in POS kiosks.

## 1.4 Research Questions

In order to execute the study meaningfully, a broad research question was developed from the study objective. The broad research question is: what is the level of awareness among POS kiosks owners about the possibility of that insider threats in POS kiosks can lead to cybersecurity issues? Four specific research questions were derived as means for providing answers to the broad research question. The specific research questions include:

- Is there a good level of awareness among owners of POS kiosk owners that ensures that they serve as guardians against insider related financial threat in POS kiosks?
- Are the POS Kiosks owners and customers aware that POS attendants can become likely offenders who perpetrate insider related financial crimes in POS kiosks?
- What kind of insider related financial crimes have POS kiosks owners experienced and how are these instances connected to the information security revolving around the business processes adopted in POS kiosks.

## 1.5 Justification of the Study

This thesis can be justified by looking at a new theme that is likely to constitute a critical challenge to the adoption of Information Technology in the Nigerian financial environment, namely insider related financial crimes in POS kiosks. This is to say that the study will provide theoretical knowledge on insider related financial crimes that may occur in POS kiosks and how the crimes can lead to cybersecurity threats. Carrying out the study is important because theoretical knowledge provide the basis for practical knowledge. In other words, the study is likely to provide theoretical insights from which important practical knowledge can be derived on how to identify and manage POS related financial crime that may lead to cybersecurity issues.

## 1.6 Conceptual Framework

The study is conceptualized around three important stakeholders that are mainly involved in the use of POS kiosks for financial services. The stakeholders are namely, POS kiosks owners, POS kiosks workers and POS kiosks customers. POS kiosks owners adopt POS to set up financial service outposts on behalf of banks and in the process plans to make a living from the profits accrued to the business. POS kiosks owners employ workers who serve as kiosks attendants and are responsible for carrying out the day-to-day business activities done within the POS kiosks. POS kiosks customers are the people who use POS kiosks either to receive money from their bank accounts (or third party accounts) or to send money, usually cash deposited to POS kiosks into other peoples' bank accounts.

The POS kiosks owners may not be on ground in the kiosks and therefore be involved with transactions involving receiving money by POS kiosks customers. POS kiosks attendants are usually on ground and are involved in all the business processes that are completed for a POS customer to receive money. Usually, the POS kiosks customers provide their ATM cards to the POS kiosks attendants. The ATM cards are slotted into the POS machine after the POS attendant might have entered required POS codes. Then the POS is handed over the POS customers to enter his/her ATM personal identification number (PIN). After entering their PINs, POS kiosks attendants collects the POS machine ensure that the transaction was completed or rejected. Every transaction, whether completed or rejected, is shown on the POS machine screen and confirmed by a printout made automatically by the POS machine.

If the transaction is completed, the POS kiosks attendant gives a copy of the printout as receipt to the POS kiosks customer and then proffers the amount of money indicated during transactions. If the transaction fails, the POS kiosks attendant provide the rejection receipt to the POS kiosks customer and could repeat the process to see if the transaction will be completed. There are many possible risk in the POS kiosks transaction. First is the revelation of personal information which has to do with POS kiosks customers' bank name, ATM card number, date of expiration of ATM card and the ATM card's CVV number. During transactions that involves the sending of money into other peoples' bank accounts, POS kiosks customer are required to write their names and the names of the persons they are sending money into their bank account. Overtime, given repeated transactions by POS kiosks customers, the amount of personal information that can be put together by POS kiosks users can be enough to provide detailed information that may jeopardize the cybersecurity of the POS kiosks customers.

Consequently, the study is conceptualized to assess and understand the extent to which POS kiosks owners are aware of the threats POS kiosks transactions poses. In the context of the study, the POS owners are taken as capable guardians given that they have the power to determine who to hire and how to design business processes in ways that may lead to insider related cybersecurity threats. POS kiosks customers are taken to be suitable targets given that they provide information relating to their bank accounts and could do this without taken enough precautions. POS kiosks attendants are taken to be likely offenders in that they have access to private financial information of both the POS kiosk owners and customers. The Diagram below shows the relationships among the variables identified in the study.

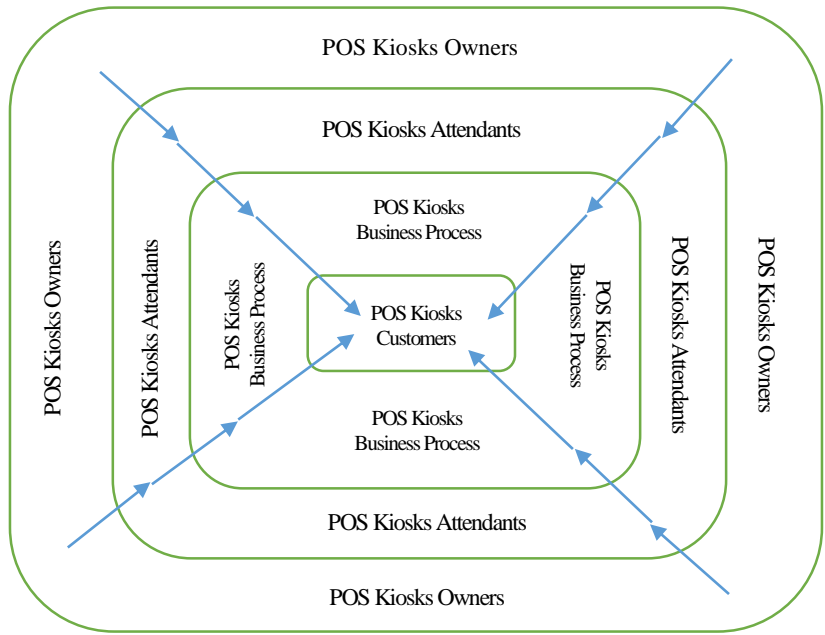

Figure 1: Key Entities in the Study

The study is conceptualized to show how the interconnection of the entities namely, POS kiosks owners, POS kiosks attendants, POS kiosks business process and POS kiosks customer may lead to financial service security threats that may evolve into cybersecurity issues. The four entities are all involved in the business processes that are completed when POS is used as means for carrying out financial services. Consequently, understanding their roles from the perspectives of capable guardian, likely offender and suitable target is likely to expose how insider related financial crimes that may be perpetrated in POS kiosks can be ameliorated.

## 1.7 Definition of Key Terms
There are four key terms in the study and they are defined below:

**POS kiosks owners**: These are individuals who own the POS kiosks. The have registered businesses whose names are used to buy the POS machines from commercial banks. They provide the finances and leadership in the POS kiosks business.

**POS kiosks attendants**: These are individuals that are hired by POS kiosks owners to see to the day-to-day running of the POS business. They operate the POS machine, initiate bank transfers and deal directly with customers who patronizes the POS kiosks for financial services.

**POS kiosks customers**: These are the individuals that make POS kiosks for financial services. They usually use their ATM cards to access funds deposited in their bank accounts using POS machines operated by POS kiosks attendants and transfer or receive money if need be.

**POS kiosks business process**: These are the business processes completed by POS kiosks attendants when helping POS kiosks customers to withdraw money from their bank accounts. The POS business process also include processes completed when helping POS kiosks customers to deposit money into other peoples' bank accounts. POS kiosks business processes also include process completed to keep records of financial transaction done in POS kiosks. The records of financial transactions are useful to POS kiosks owners and customers in cases where there are discrepancies and are also used as proves of transactions in case referred to commercial banks.

# CHAPTER TWO

# LITERATURE REVIEW

## 2.1 Financial Services and Information Technology

Every society has several modes in which financial services are provided and the security of the processes involved are well placed and in order Records have shown that there were various ways different traditional societies in Nigeria dealt with financial services and the adjoining financial service security required for the system to run well. There were stories about the period when people put their legal tender (cowries, gold, money and other forms of legal tender) in local pots and hid them in the ground. There are also stories about when shrines and religious worship centers served as the place where people kept their legal tender (Dam, 1981). During this era the priest assumed the role of financial service custodian and helped the people create a financial service systems that had enough integrity to ensure the security of the funds. All the gestures that were enacted during the traditional times were promoted by the need for accountability and safe keeping of money during the epoch. Although there were very little stories about the insecurity of traditional financial service systems, those who lived during the epoch however, had different ways for safeguarding and ensuring the safety of their financial service systems. The gestures also meant that they harbored the fear of the financial systems at the time can be prone to insecurity (Dillard & Vinnari, 2019).

The spread of international businesses mainly from Europe into Nigeria, and the business transactions that started to occur between Nigerians and European business owners long before colonialism laid the foundation for the evolution of money as legal tenders in Nigeria (Davies, 2010; Falola & Heaton, 2008). It was however, colonialism that resulted to the use of money and establishment of the first sets of banks in Nigeria during the colonial era. The growth of civil service and increase in the number of private businesses, including, foreign owned business necessitated the establishment of banks in Nigeria. Historically, the Bank of British West Africa which was established in Nigeria in 1894 was the first commercial bank in Nigeria (Falola & Heaton, 2008). Although the bank was established in Lagos, Nigeria it was meant to service the financial product and service needs of all the British Colonies in West Africa. The colonies include Ghana, Cameroun, Sierra Leone and the Gambia. The Bank of British West Africa, Lagos, Nigeria was acquired in 1965 was acquired by Standard Bank, Johannesburg, South Africa and in 1979 it was localized and became First Bank of Nigeria.

In Nigeria, like in other places across the world, commercial banks provide bank products and services that can be categorized into two broad types namely, corporate bank products and services and individual bank products and services. The corporate bank products and services are those that are targeted at corporate organizations including private businesses, public businesses and government agencies. Individual bank products and services are rendered to individuals who are constitutionally eligible to bank products and services (Efobi et al. 2014).

The invention of information technology (IT) resulted to the discharge of all the commercial bank products and services using IT. The use of IT for banking services started with the deployment of websites used by banks to provide information about themselves, and products and services. At the initial stage commercial banks' websites were static websites (Zolait et. Al. 2008). This served the purpose of advertising and for providing information about bank products and services. The improvement in web technology resulted to the use of commercial banks' websites for actual transaction of banking services (Khan, 2018). It also resulted to the use of portals for database management systems that commercial banks used. The portal systems used by commercial banks laid the foundation for the automation of banking transactions.

Consequently, the evolution and popularization of the electronic banking systems was promoted by the vast improvements in the capabilities of web technologies to accommodate interactive features. Electronic banking, otherwise known as the e-banking systems, encompasses the use of electronic platforms which are made possible by the integration of computer technology, internet technology and software technology for transacting banking business processes that enables commercial banks and their customers to transact business without meeting face-to-face (Inegbedion, 2018). E-banking systems radicalized deposits, withdrawal, payments, remittances, and the management of foreign exchange transactions. At the initial stage, customers completed processes required to access, utilize these products and services using commercial banks' websites. And they normally do the transactions outside the banking premises and unaided by staff of commercial banks. The issue of cybersecurity, commercial banking and commercial bank customers resulted in the wake of the pertinent and essential use of IT for banking operations and services.

## 2.2. Mobile Financial Services

The influx of mobile devices and the promises this technological device offers for everyday needs resulted into the evolution of mobile financial services. Mobile technologies and devices helped commercial banks to dramatically reduce the limitations connected to the adoption of e-banking. This is given that, at the initial stage, it requires users to own internet complaint personal computers and to have access to the internet. Most of the IT devices available at the initial time when e-banking was introduced were heavy and could not be moved around. This limited the possibility of real time usage and for everyday activities whilst an individual is at home, work or a recreational setting requiring financial use. It also resulted into a situation in which e-banking was seen as an official issue; issue that must be completed in formal setups, particularly in offices. Mobile devices, particularly mobile phones, radicalized the e-banking systems. Given the ubiquitous nature of mobile phones, its low cost and easy-to-use nature, they provide a cheaper and more convenient platform for assessing commercial banks' products and services. All the limitations connected to the use of e-banking before the advent of mobile devices were solved by the introduction of the mobile devices. For instance, mobile devices are not heavy and could be carried around by their owners. It also provided avenue for real-time usage. Mobile devices brought about the perception among users, that e-banking is an everyday life issue, unofficial and can be completed anywhere and at any time.

The evolution of the term mobile (m) banking and its fast replacement of the term e-banking was as a result of the popularity in the use of mobile devices to access financial products and services that were accessed using IT devices. Corbitt (2003) posits that the convergence of the internet and mobile technology and the use of the convergence for access banking products and services resulted to the evolution of the m-banking. Accordingly, Barnes & Corbitt (2003) argues that,

> "In the last decade, we have witnessed the increasing penetration and development of two key areas of technology: the internet and wireless telephony. The internet has proven to be an easy and efficient way of delivering a wide variety of services to millions of 'wired' users; as of February 2002, the estimated number of internet users stood at 544 million, rising to 1 billion by 2005 (p. 2)."

There is an agreement amongst scholars that m-banking is unusual and that it provides opportunities that have never been provided for by other technologies that existed before it. Consequently, there have been studies that were carried out to assess the benefits of m-banking. Herzberg (2003) for instance, declared the advantages of m-banking over other types of banking strategies including the popular e-banking. He argues that mobile "…devices are effective for authorizing and managing payment and banking transactions, offering security and convenience advantages compared to online payment via PCs. Some of these advantages are available in existing devices, others require modest, inexpensive enhancements likely to be available in new devices in the next few years (p. 53)."

A major factor about the usefulness and advantages of m-banking is that every stakeholder involved in the provision of banking products and services have advantages they derive from it. For instance, m-banking is a very powerful automation system that enables regulatory bodies that work with banks with respect to assessing banks against regulatory requirements easily. One important area regulatory organizations benefit from when it comes to regulating banks activities is the quality and integrity of operations records. Catherwood (2011) for instance, argues that m-banking is a very useful for the detection of money laundering. In a more recent study, Whisker & Lokanan (2019) present how m-banking could help regulators and governments to avoid money laundering and terrorists' financing networks. The issue of money laundering is a global phenomenon that has posed several challenges to governments, individuals and organizations. Money laundering has to do with changing money obtained from illegal and criminal sources into forms that indicate that such money was legally obtained. Governments, individuals and organizations including for-profit and not-for-profit organizations have suffered severely from money laundering. The new banking designs and implementation of their products and service operations are usually automated and are largely influenced by m-banking requirements. M-banking requirements provide avenues for regulators to use bank transactions records systems to track and monitor funds deposit, transfers and withdrawals.

Aside the crime aspect of the benefits of m-banking, there are also many ways m-banking has helped regulators to initiate, implement and monitor development initiatives. The financial inclusion initiative is one of the many development programs that were initiated given the dynamic nature of m-banking. According to Olagunju & Utulu (2021), financial inclusion has to do with creating equal opportunity for people, irrespective of their gender, creed and tribe, to have access to funding making ventures and financial products and services banks and other financial institutions provide. Given that financial inclusion challenges are more profound in rural and poor contexts; mobile devices have been found to be very useful to achieving set financial inclusion goals. Many financial services have been made available in rural setups because of the possibility to use mobile devices, particularly mobiles phones, to provide financial services (Ouma et al. 2017). The growth and the ease with which international migration is supporting economic growth in developing countries through remittances has been acknowledged as a positive development. Mobile devices and the possibility to use to the access financial services have had very dramatic impact on remittances and poverty reduction (Wagle & Devkota, 2018). Given the dynamic nature of mobile devices and the financial services that have been developed by banks due to this, key stakeholders do not exercise any doubt that mobile devices have tremendously helped to improve on access to, and use of financial services.

The evolution of POS and the advantages it offers to the delivery of financial services is a very good example of the impact of mobile devices in the delivery of financial services. POS is useful in every context. Big one-stop departmental stores use POS to facilitate payments by customers (Mafimisebi et al. 2019). Small and medium scales enterprises also use POS to facilitate payments made by customers (Olasojumi et al. 2018). In every society including those that are tagged developed and developing, POS has been found to be a very applicable and reliable payment management system. Individuals that render services use POS to collect money due to the services they render. Organizations also use POS to collect and make payments. The cashless policy initiative in many developed and under-developed countries have been supported by the adoption of POS (Utulu, 2021).

## 2.3 POS Financial Services and Cybersecurity

There are many good things about the use of POS by individuals and businesses for financial services. However, there are also many challenges that are connected to the use of POS by individuals and businesses. In the recent past, scholars and other stakeholders started to develop keen interests in cybersecurity issues. The cybersecurity issues also have connections to the use of POS by individuals and businesses. The use of POS exactly fits into cybersecurity issues because they are used with the aid of the internet. Consequently, most cybersecurity issues that constitute concerns to individuals and businesses that use the internet with other types of hardware also give concerns to those that use POS. In fact, studies carried out in the recent past has shown that the possibility of cyberattack is one of the major challenges of using the POS (Hassan & Utulu, 2022). In Nigeria for example, the frequency with which cyberattacks occur is unprecedented, rampant and widespread. The condition is more worrisome in states that are experiencing armed conflicts and those close to them where people move to as internally displaced people. Consequently, there are studies that reveal that cyberattack is one of the major threats to the use of POS in Nigeria (Hassan & Utulu, 2022; Onimisi & Noyenlum, 2018). In fact, cyberattack concerns have been reported as one the major threats to the adoption of every form of IT based financial service strategies in Nigeria (Igoni et al. 2020). Cybersecurity concerns that impact on the effective use of POS take many shapes and forms. They are issues that have to do with insider threats where an employee of a financial institution supports attackers' efforts to gain illegal access to financial records. There are also cases of careless by owners who may due to personal or social reasons give away their personal security information and in process enable attackers to access their financial cyberspace.

Another cybersecurity issue that is linked to the use of POS is the giving away of sensitive personal data. POS use requires the use of ATM cards with inscribed personal information such as names and personalized information such as card number, card verification value (CVV) number and date of expiry of ATM cards. The way ATM cards are designed provides easy access to information that are vital to its owners and for its use. The information on ATM cards is critical to making transactions on POS and other IT based financial transaction systems. Paradoxically, the information can be easily cloned and stolen. (Kaur et al. 2019). Scholars have made good efforts to study the shapes and the factors that promote the breach of cybersecurity that are put in place to avoid ATM related cybersecurity issues. Kaur et al. (2019) focused on cloning and social issues social developing adequate and appropriate ethical regulations that could help combat ATM related cybersecurity issues. Hattali et al. (2020) proposed the design and implementation of an automated ATM fraud detection and prevention. Hattali's et al. (2020) study is a pure state of looking at ATM related cybersecurity issues from the technical perspective where the cause and solution are dependent on the technical capabilities of the stakeholders.

There is no doubt that cards pose a lot challenges to both organizations and individuals. Despite this however, it seems that stakeholders are not properly looking at the challenges that ATM cards pose in POS kiosks. There are many reasons why stakeholders should worry about the cybersecurity challenges ATM cards pose when it comes to using them in POS kiosks. First, is the history of insider threats in formal financial service institutions like commercial banks and related banks. The literature on insider threat in commercial banks in Nigeria is fast growing and shows that the threat is becoming worse by the day (Tade, 2021; Samuel et al. 2018; Roy & Prabhakaran, 2022). In POS kiosks, the owners and personnel do not have formal training in financial management. The idea behind implementing POS kiosks is to improve on the rate of access to financial services among customers who may not be close to a bank. Often times, the owners of POS kiosks employ people without formal integrity checks. Integrity check on individuals is complex and requires a good level of profiling. The use of profiling to ensure that the personnel hired to manage POS kiosks has not been studied. So, questions regarding if POS kiosks owners profile the personnel they hire has not been addressed by scholars.

Invariably, POS kiosks owners are likely to employ people they know, which refers to people that are socially and physically close to them. It is highly likely that they employ family members, friends or people that they are familiar with in the neighborhoods where they live. These are expected social factors that are likely to determine those employed by POS kiosks owners and could pose an insider threat to the cybersecurity of POS kiosks. Trust and familiarity may result among POS kiosks owners and the people they employ and could result in situations where the required integrity checks are not done. The lack of integrity checks poses great danger that can easily lead to insider threat in POS kiosks. Although insider threats can easily result from hiring a corrupt person that was not subjected to integrity checks, the business processes of POS kiosks also provide good avenues for cloning ATM cards. Usually, customers who use their ATM cards to transact businesses in POS kiosks submit their ATM cards to operators who are seated in POS kiosks. POS kiosks are usually small and can only accommodate the operator. They are also covered for physical security reasons. Transactions involving the use of POS machines and customers' ATM cards are usually done within the kiosks and this makes it difficult for the customer to know what the POS operator is doing with his/her card. The business processes completed in POS kiosks during financial transactions are designed in ways that exclude customers from knowing what the POS operator is doing with their ATM cards.

Consequently, there is need to reassess those hired to work as POS operators, the kind of integrity checks POS kiosks owner carry out to ensure that they do not pose cybersecurity threats and the business processes completed when a customer uses POS kiosks for financial transactions. These issues still do not form part of the issues that scholars focus on as evident in the literature. There ae many research studies that are focused on cybersecurity issues that may occur when ATM cards are used, many of the studies do not look into ATM cards uses, cybersecurity and POS kiosks.

## 2.4 Theoretical Framework

Theories serve as frameworks for carrying research in the social sciences including, in the information systems field. Theories provide the basis for the kind of theories a research is likely to produce with his/her study (Gregor, 2006). It also provides grounds for the factors that the researcher is likely going to focus on. The victimization theory provides the theoretical grounds for the narratives sought for in the study with regards to how the three stakeholders-POS kiosk owner, customers and attendant-may be involved in the social construction of the realities that may lead to insider related financial threats in POS kiosks (Choi, 2008). The victimization theory argues that the characteristics of the victim precipitate crimes (Choi, 2008; Hindeland et al. 1978). The victimization theory is divided into four major theories namely, victim precipitation theory, lifestyle theory, deviant place theory, and routine activity theory. The study adopted the routine activities theory.

### 2.4.1 Routine Activity Theory

The routine activity theory is a sub-set of the crime opportunity theory. The crime opportunity theory evolved in the field of human security and proposes that crimes occur in social situations. It argues that the social situations are characterized by opportunities that are either deliberately or not deliberately provided to criminals. It further explains that criminals only commit crimes when they have assessed the opportunities for crime and ensured that minimal efforts are required to commit the crime without being caught (Back et al. 2021; Wilcox, et al. 2018). In real life situations, the opportunity to commit crimes by criminals are usually given by people without knowing. It follows that most times, people ways of life and actions exposes them to make decisions and takes actions that exposes them to criminals. This inclination is at the center of the routine activity theory in that it proposes that social change and the kind of behavioral changes that evolves as a result provide criminals with opportunities to commit crimes (Cohen & Felson, 1979). In the recent past, many studies have been carried out that were informed by the notions propagated in the routine activity theory. This is given the upsurge of different forms of social changes, including those that were occasioned by Corona Virus. For instance, Halford et al. (2020) studied how corona virus induced social change including social distancing, lockdown, etc. triggered new criminal behavior and crime rates.

This study uses the routine activity theory given that it was observed that the evolution of POS kiosks has resulted to the evolution of some social behaviors that may provide avenue for criminals to steal financial information that are crested on ATM cards. The ways POS kiosks are designed, the business processes that are followed by those running them, and the extent to which POS kiosks owners vet the integrity of the people they hire constitute new social realities that may trigger the occurrence of new form of ATM card cloning and cybercrime. The study pays attention to POS kiosks staff as likely offender given that little is done to check how POS kiosks business processes can enable them commit and/or aid those that want to commit ATM card cloning and cyber related financial crimes. The study also takes POS kiosks owners and customers that transacts financial businesses using POS kiosks as suitable targets. It was also taken in the study that POS kiosks owners and scholars the capable guardians. This is given the role of the POS kiosks owners as the strategy developer and strategy implementation enforcer. Scholars are taken as capable guardians because they are expected to push scientific inquiry into how POS kiosks may expose stakeholders to new forms of financial service related crimes.

# CHAPTER THREE
# RESEARCH METHODOLOGY

## 3.1 Research Design

Consequently, the study adopted the qualitative research design to assess POS kiosks owners' awareness level of the possibility of insider related financial crime which may lead to more serious cybersecurity crimes. Issues relating to insider related financial crimes in POS kiosks have not enjoyed the attention of researchers despite the growing number of POS kiosks across most African countries including Nigeria. There is also a lack of awareness among stakeholders, particularly scholars that study cybersecurity issues related to using IT for financial services, that the use of POS kiosks can lead to the creation of new gateways for IT related financial crimes. These two gaps in existing scientific knowledge requires a firsthand information that could be gather in a qualitative design based small scale study. This is the reason why the study adopted the qualitative design.

Alternatively, there are two other forms of research designs namely, quantitative and mixed method design that could have been adopted to execute the study (Turner et al. 2017; Watson, 2015).. The quantitative research design supports large scale studies which are carried out to have a holistic view of the subject(s) that the researcher aims to study. Both the quantitative and mixed method research design were not found to be applicable to the study. This is because the issue under investigation in the study is still novel and requires firsthand information before a large scale study could be carried out more effectively.

## 3.2 Research Approach

The research approach used in study is the deductive research approach. The deductive research approach provides the ground for researchers to guide their investigation with existing knowledge (Johnson-Laird, 1999). The existing knowledge are usually in the form of theories that have been published by other scholars or the same scholar. The logic behind the deductive research approach requires that researchers do the investigation of phenomenon following a thinking that enables them to use specific knowledge about the phenomenon to generate a general knowledge about it (Hyde, 2000). This is to say that the researcher starts his investigation into a phenomenon by using a specific knowledge which is usually a theory. In the case of this study, the routine activity theory provides the specific view about the phenomena, financial crimes in POS kiosks, that was investigated. In other words, the ideas used to execute the study were derived from the routine activity theory. The initial specific knowledge derived from the routine activity theory was used to come up with the general knowledge that were used to come up with research model presented in the study. Other research study approaches that could have been used in the study are the inductive approach, retroductive approach and abductive approach.

## 3.3 Research Philosophy

The interpretive research philosophy was adopted in the study. Studies that are devoted to investigating social issues usually adopt one of the available four research philosophies. The research philosophies are namely, the interpretive philosophy, positivist philosophy, critical realism and pragmatist philosophy (Alharahsheh & Pius, 2020). The central argument in the research philosophies revolves around whether social phenomena exist naturally or based on the outcomes of human thinking and actions. While the interpretivist philosophy argues that social phenomena are products of human and actions, the positivist philosophy argues that social phenomena are products of nature. To the positivist philosophy, social phenomena are natural and objective phenomena (Alharahsheh & Pius, 2020). Whereas, the interpretivist philosophy sees social phenomena as subjective and a creation of human actors that are involved in the enactment of the phenomena. The critical realism philosophy takes a central stance and argues that social phenomena come into existence objectively, but that human actors enacting the social phenomena are the ones that subjectively interpret them. This therefore means that critical realism philosophy promotes the notion that existence is both within the realm of objectivism and subjectivism.

Consequently, the research philosophy adopted in the study is the interpretivist philosophy. This is to say that it is believed in the study that social phenomena are subjective and created by the social actors that enact them (Utulu & Ngwenyama, 2017). The implication of this to the study is that every entity identified and dealt with in study are taken to have subjective existence, that is, they are all created by those that enact the social actions surrounding them. For instance, the POS, POS kiosks, POS kiosk owners, POS kiosks attendants and POS kiosk users are all involved in the creation of the realities surrounding the financial crimes that may occur in POS kiosks. This is to say that all identified entities are created by social actors.

### 3.4 Research Method

The case study research method was adopted in the study. The case study research method promotes a study strategy that enables in-depth assessment of a small number of examples, usually called cases (Flyvbjerb, 2011). The case study research method could be in the form of qualitative case study research method, quantitative case study research method or mixed method case study research method. The variation of the case study research method that was adopted in this study is the quantitative case study research method. It is quantitative case study research method because it collected and used the quantitative data as the basis for reaching the conclusion reached in the study (Gerring, 2008). As discussed earlier, the study is a small scaled study and seeks to provide firsthand information on phenomena surrounding the extent to which POS owners are aware of the possibility of insider related crimes in POS kiosks. It therefore suits the provisions of the case study research method. As a result, the study chose a small number of POS kiosks to serve as the unit of study in the case study research method.

**3.5 Study Population**

The population of a study in studies investigation social phenomena comprises the total number of the entity that shares unique characteristics that distingue them from other entities. In the study, POS kiosk owners and POS kiosks are the entities under investigation. POS kiosks owners and POS kiosks possess unique characteristics that distinguish them from other entities that are connected to financial services. For instance, POS kiosk owners are different from owners of other financial service outlets let, for example, microfinance bank owners. So also, POS kiosks have unique characteristics that are different from those of other financial service outlets, for instance, microfinance banks and POS financial services done on desks place in open spaces. Having explained this, the study population comprises of the entire population of two entities namely, POS kiosks owners and POS kiosks. This is to say that the study population comprises of POS owners and POS kiosks in Yola.

**3.6 Study Sampling Technique and Sample Size**

Sampling techniques are used to select a fraction of the study population that are studied by researchers. Because most studies compromises of study population that are large to study, sampling techniques were developed to enable researchers to select the required proportion that are actually studied. Consequently, the sampling technique used in the study is the purposive sampling technique. Purposive sampling technique is one the sampling techniques that are appropriate for selecting study samples in research studies that are based on qualitative design research. Purposive sampling technique therefore do not use statistical rules to determine the sample that are fit and adequate constitute a study's sample size.

The purposive sampling technique is subjective and based the primary parameter used to select a sample on the purpose of the study and the sample that actual fits the purpose of the study. In study, given that the phenomena that is of interest is the assessment of POS kiosks owners' awareness of possible insider related financial crimes, the study sampled people that own POS kiosks purposively. This is to say that they are sampled because they match the exact purpose of the study namely, assessing POS owners' awareness of possible insider related financial crimes in POS kiosks. The study did not use any statistical rule to determine the POS owner that was appropriate to be studied. POS owners that were chosen were chosen based on the fact that they own POS kiosks. 15 POS owners' and by effect twenty POS kiosks constitute the study sample size.

## 3.7 Data Collection Technique

As indicated earlier in the study, the qualitative data was collected to serve as a basis for reaching conclusion in the study. There are several techniques that can be used to collect qualitative data including, interviews, observations and secondary data in document forms. The interview method was used as the data collection technique in the study. The interview data collection technique also can be in the form of structured interviews, semi-structured interviews and unstructured interviews. However, the study adopted the semi-structured interview technique. This is to say that the interview questions were derived from both the routine activity theory and insights that are derived during the course of the research. The choice for semi-structured interview was to extract sensitive information, it also falls in line with preparing interview questions before the interview and deriving data for this thesis. In other to initiate the semi-structured interview, ten tentative interview questions were raised as guide questions from insights in the routine activity theory. Other interview questions that were used in the study emerged based on discussions held with study participants which are referred to as the POS owners. POS kiosks were interviewed to get their views on insider related financial crimes in POS kiosks. The interviews were recorded with the recording device attached to a mobile phone and the interview questions are

Demography: Questions posed to POS Kiosk Owners

1. How long have you been doing your POS business?

2. How many persons have you hired

3. How do you know the right person to hire?

4. Why do your staff leave?

5. Have you ever sacked any staff? And why?

6. Are you aware of the risks attached to the POS kiosks business

7. Do you think your staff pose a challenge?

8. The processes used to complete transactions, do you think there a risks associated with it?

9. Have you ever thought about these risks before now?

10. In all how do you ensure your security and that of your customers?

### 3.8 Data Analysis Technique

The thematic data analysis technique was used as the data analysis technique in the study. The thematic data analysis technique is based on identifying similar themes within interviews. The similar themes are usually determined by subject headings that may have been derived from the variables available in theories. In the study, the routine activity theory provided the subject headings that were considered as the themes used to group the interviews into the variables that indicate the level of POS owners' awareness of possible insider related financial crimes in POS kiosks. Themes were derived from the interviews through the following processes. The recording interviews were transcribed into textual form and printed out. The interview transcript printout was read many times for familiarity sake. During the reading of the interview transcript similar themes relating to identified financial crime variables were grouped into themes and were used to provide explanations on the study findings. These steps are based on Braun & Clarke's (2019) six-step procedures for carrying out thematic analysis and they include, familiarizing oneself with the data, generating initial ideas, looking for themes, reviewing themes, defining and labelling themes, and publishing the report. Appendix 1 contains example of empirical evidence supporting the key empirical observations in the study

### 3.9 Ethical Issues and Declaration

Research studies that involves human subjects normally require meeting laid down ethical rules. The ethical rules are required in order to avoid the violation of the rights of the human subjects which usually revolve around voluntary participation and non-disclosure of sensitive personal information. The study was carried out therefore after meeting some ethical requirements that are both at the individual level and institutional level. First the research study was registered as a research study required in partial fulfilment of the award of the bachelor degree in information systems. The supervisor of the project has a considerably good level of research experience. The supervisor ensured that all institutional requirements required to be met before starting a research study that will engage human subjects are meet. At the individual level, all human subjects namely, POS kiosks owners that participated in the study were introduced to the requirements of the study before they were formally engaged in the study.

Their consent to voluntarily participate in the study was sought. Research participation consent form was filled by them and they were assured that they can decide to withdraw their participation in the study at will. There was also a formal agreement that information provided in the course of the study will be treated confidentially and that they will be used only for the purpose of the study. The study participants also had the power to decide if the interviews held with them should be recorded or not. The interviews were held in a way that ensured that personal information such as names, name of business, location of business and business banker was not revealed and/or recorded.

# CHAPTER FOUR

# PRESENTATION AND DISCUSSION STUDY FINDINGS

## 4.1 Presentation of Study Findings

The study's objective is to assess the level of awareness among POS kiosks owners about the possibility that insider threats in POS kiosks can lead to cybersecurity issues. The study was necessitated by the growing number of POS kiosks and those that patronize them. To reach the objective of the study, first hand data were collected and the analysis shown below. Fifteen POS kiosks were studied. Interestingly, all the POS kiosks are owned by men. Majority of the POS studies have been in existence for about two and more years and are managed by a POS kiosk attendant. The POS kiosks make considerable amount of money pay day through payments and transfers.

According to the POS kiosks owners, the primary financial service they render is payment services. In other words, they provide payments in the form of withdrawal to people. Most users of the POS kiosks are those who for several reasons may not want to use an ATM machine in bank premises. Another financial service POS kiosks render is the third person money transfer service. This service requires that people who may want to transfer money to someone else come to the POS kiosks with the amount of money they want to transfer in cash, make payments to the POS kiosk attendant who will then transfer the money to the person whose account details was provided. On few occasion however, POS kiosks also help their customers to receive money. Such money are transferred to the POS kiosks' bank accounts and cash provided to the person on behalf of whom they received the transfer.

The POS kiosks charges one hundred naira for each of these types of financial services per five thousand naira. In other words, transactions below but not more than five thousand naira are done with a service charge of one hundred naira. This also means that three-hundred-naira transaction fee will be charged, for instance, for transactions that are up to fifteen thousand naira. All the POS kiosks owners agreed that the charges for financial transactions are the benefits that they derive from the business and that they could make up to ten thousand naira in a day. POS owner. It follows that POS kiosks are good example of job generating IT initiatives in the financial service sector.

Consequently, most of the POS owners are young males and have some level of formal education. All of them could speak good English Language. Three of them indicated that they have POS kiosks in different locations. Out of the three POS owners that have multiple POS kiosks, two have two POS kiosks located in different locations in the Yola Metropolis, while one of them has three POS kiosks located in different locations in the Yola metropolis. All the POS kiosks studied have one POS kiosk attendance who carries out the day to day running of the financial service business for the POS kiosks' owners. All the POS kiosks also have a ledger, usually a higher education note book with hard back. The ledger serves as the records system for recording details of people that carried out transactions that require documentation. Usually, documentation is done for transactions that have to do with sending out money on behalf of a third party. The third party writes his name, account number where he wants the money to be sent, the same of the owner of the account where the money is to be sent, mobile phone number and date. POS owners revealed that the data is useful for tracing transactions in cases where issues like failed transactions arises.

The study data shows that all the POS kiosks owners are not aware of the possibility for an insider related financial crime that could evolved into cybersecurity threat. This is quite surprising given that they are aware of a number of financial crimes that have been perpetrated in POS kiosks. POS kiosk owner I revealed that *"I never thought about that and it looks like it is important to* *think about such things…all I think about is how to avoid customers to cheat us either with fake alert or use of magic to take our money."* The issue of fake alert seems to be very crucial and popular in the Yola metropolis. It occurs when a POS kiosk user uses bank transfer services, usually the USSD based bank transfer services to send money into POS kiosks' account in exchange for cash. The use of magic to take money is very peculiar to the Yola metropolis as indicated by the POS kiosk owners. The crime occurs when a customer uses magical money to transaction financial business with the POS kiosks. It is believed that once the money with magic is put among the money in the POS kiosks that the 'real' owned by the POS kiosks disappears, allegedly to the culprit's pocket. POS owner X revealed that, *"every POS kiosk owner warns his staff about this. We tell them to look very well at the people that come to patronize them and that they should not put the money received from such people with the money we already have."* POS owner VII also made comments on it: *"the problem of some traditionalist moving around with magic money to take our money away is very serious. It happens in the market and can also happen with us. Although people who this, target big money making business where they can steal millions."* With regards to fake alert, POS owner VII reveals that, *"I pray I should not experience it. I have heard about it even before I started my POS business."*

Consequently, POS kiosks owners' attention was fixed on financial crimes that could be perpetrated by others against them and the insider kind of financial crimes. This is despite that they all agreed that the ATM card and the POS mobile machine contain crucial information that could be used to carry out unauthorized businesses. With regards to the information inscribed on ATM cards, POS owner II testifies that, *"It is troubling that I cannot think about other ways ATM cards can be produced to reduce the open information it carries. It just too easy to steal information on ATM cards. So owners must be careful and make sure they do not fall victims"* POS owner VI argues that, *"the ATM card is risk anyway…but being careful is the only way to ensure that others do not copy your information. For me, I am very careful with mine and I encourage others to."* Despite this, majority of the POS owners agreed that the ways POS kiosks are designed and the ways to run their business can enable insider related threat to information on ATM cards. POS owner IX posited that *"I think that you are right. The whole problem just surfaced in my head now. If one is not careful your POS boy can actually take someone information and use it whenever he wants to. This means that customers too must request that some protocols to ensure that their cards are used in on the desks so that they can see what's going on."* The questions raised during the data collection on the possibility of insider related threats that could lead to other forms of cybersecurity threats seem to serve as eye-opener for the POS kiosks owners. Although before going to the field, it was difficult for me to assume that POS owners never thought about the possibility. It therefore follows that there is a low level of awareness among POS owners that an insider threat that could lead to financial crimes that are likely to evolve into cybersecurity crimes could occur in the kiosks.

Given this therefore, it was further probed if they actually do any form of profiling to the attendants that they hire as a way of ensuring that they do not pose any form of financial security threat. On this, all the POS owners indicated that they do not just hire any person to serve as their attendants. They noted that they hire those that they know directly or people that they know their family members. The also indicated that they go to where the person they want to hire lives and ask questions about his character from people that know him. POS owner I indicated that *"I had to go the street where my boy, the one here, lives and ask about him before I hired him. The other shop, the boy there lives in my street and I know him very well. So I didn't have any doubts about his character and integrity. In the third shop, the boy in my second shop who lives close to my* *house is the one that brought that one. He is a good boy…at least my boy told me he cannot run away with my money."* It follows that the type of profiling that POS owners do is still basic and depend largely on social relationship. This is expected because the Yola-South metropolis is not as sophisticated as other societies outside Nigeria where official records of individuals can be easily accessed. The ways POS kiosk financial service business are perceived by their owners come to play in the ways the attendants that POS kiosk owners hire are profiled for character confirmation. First, the POS kiosk owners have never talked about insiders' threat before the study. Second, the POS kiosks owners also only think about frauds perpetrated by other people against them as the only possible crime to look out for. Consequently, when profiling the attendants they hire, what they have at the center of their mind is the person who will find it difficult to run away with their monies. It follows that in its present state, POS kiosks are still prone to insiders' related financial crimes that could lead to cybersecurity threats.

## 4.2 Discussion of Study Findings

There are lots of research that have been carried out and published on POS. The POS happens to be an integral part of the evolving mobile banking (m-banking) evolution in the twenty first century. The primary factor, as documented in the literature that promotes the use of the POS at the initial stage of its invention is the need for efficient and effective payment systems for retail businesses (Willems et al. 2017; Banerjee & Banerjee, 2000). Over time POS evolved to become a socio-economic development promoting financial technology that were used to transact different financial services for fees (Nandonde, 2018). Unexpectedly, the use of POS for transacting financial services in kiosks has not been looked into properly by researchers. Available studies on the contribution of POS to socio-economic development is still limited to those that looked at their uses by small and medium sized businesses, particularly retail businesses (Ogunsuyi & Tejumade, 2021). Studies that looked into how POS helps individuals to meet their financial service needs have also been produced in the literature. These studies are mainly related to those that looked into the use of POS to promote the financial inclusion of people living in areas that are disadvantaged when it comes to access to financial services (Lee et al. 2020).

An important area that has been fully neglected by researchers is the use of POS for financial services in kiosks. This is despite that the number of POS kiosks in existence in contexts where there is shortage of bank financial services is growing exponentially. POS kiosks are kiosks which provide financial services such as money withdrawal, transfer and collection to people for an affordable fee. Scholars seem not to show interest in studying issues relating to the social factors that may impact on the use of POS kiosks given the role expected to be played by their owners as capable guardians, attendants as likely offender and customers as suitable targets. Also very important is that researchers have not shown interest in studying financial crime issues that may result in POS kiosks, particularly financial crime that may evolve into cybersecurity related financial crimes. This is despite that there are many studies in the literature that adopted the routine activity theory to expose and explore how social actors involved with social phenomena can constitute capable guardians, likely offenders and suitable targets (Ryens et al. 2011; Pratt et al. 2010).

Findings in this study corroborates claims in the literature about the role capable guardian could play in ensuring that the likelihood that crime with occur is reduced. POS kiosks owners were found to be very crucial to ensuring that the business processes they allow POS attendants to adopt are security friendly. Issues relating to how to play the role of capable guardian have been discussed extensively in the literature (Hollis et al. 2013; Reynald, 2010). In the study, one factor that was revealed as a determinant factor is the awareness of the POS kiosk owners that they can serve as capable guardian to both POS kiosk attendants and POS kiosk customers. This is an addition to the body of knowledge about the factors that determine if a person will serve as capable guardian or not. Similarly, there studies that have been done about the factors that may lead to people becoming suitable targets. Leukfeldt (2014) opines that routines and activities one is constantly involved in makes him predictable and therefore a suitable target. Bossler & Holt (2009) put forward that both activities carried out over time and the absence of capable guardian are likely to make a person suitable target. Findings in this study corroborates findings in these two studies. The implication is that an individual may become a suitable target because of his own activities and also because of the absence of a capable guardian. This was revealed in this study and shows that POS kiosk customers need not only pay attention to what is going on in the kiosks with regards to the security of transactions, but need to be aided by POS kiosks owners.

The important thing about this study is that it reveals salient issues that have been overlooked for a longtime with regards to ensure that financial crimes are not committed in POS kiosks. It places the financial security of POS kiosks in the hands of all three stakeholders involved namely, owners, attendants and customers. The revelations in study have implication for policy makers who may want to consider coming up with policies to ensure that financial security threats in POS kiosks are reduced to the minimum. The revelations also have implication on the POS kiosk owners in terms of the business processes they use and the extent they serve as capable guardians to customers and supervise attendants to ensure that security issues do not come up.

# CHAPTER FIVE

# SUMMARY, CONCLUSION AND RECOMMENDATIONS

## 5.1 SUMMARY

The study was carried out to look into possible financial crimes in POS kiosks in Yola, Adamawa State capital. The increase in the number of POS kiosks and those that patronize them provided grounds for the study. The assumption that the increase in the number of POS kiosks and those that patronize them may also mean an increase in the possibility of financial crimes in POS kiosks also necessitated the study. The study therefore was designed to access the awareness of POS kiosk owners with regards to possible insider threats to financial transaction security and their awareness of the fact that they can act as capable guardians. It was assumed that if POS kiosk are able to reduce the likelihood that insider related financial crimes are perpetrated in them, that this will also aid the reduction in the possibility of cybersecurity related financial crimes. This is because ATM cards carry personal information that can be stolen at POS kiosks and used to perpetrate further cybersecurity related financial crimes in more established financial institutions.

The findings of the study show that owners of POS kiosks do not have adequate level of awareness that they could serve as capable guardian. Although it was revealed in the study that they make efforts to enforce measure that can help them avoid financial crimes in POS kiosks, the efforts are primary motivated by the need to safeguard their business interest and not the security of customers. For instance, it was revealed that they do some form of profiling for those they employ as POS attendants. They also advice the attendants on how to ensure that they do not fall victim of financial crimes that will affect their businesses. They however, realize during the cause of the study that they can adopt business processes that will ensure that customers' financial information and personal data are secured. The study however, contributes to understanding how businesses processes, POS kiosk owners' role as capable guardians, customers' likely role as suitable targets and attendants' role as likely offenders can be handled.

## 5.2 CONCLUSION

The study concludes that efforts made to ensure that POS kiosks are not prone to financial security issues that may arise due to insider threats can only be effective if the role of three actors are considered. This is to say that POS kiosk owners, POS kiosk attendants and POS kiosk customers have critical roles to play in the bid to ensure that the security of POS kiosks are ensured.

## 5.3 RECOMMENDATIONS

It is recommended that more efforts should be made to educate POS kiosk owners on their role as capable guardians. It is also recommended that efforts should be made to educate POS kiosk customers on how they could degenerate into suitable targets due to their routines and behaviors when using POS kiosks. The study also recommends that more efforts should be paid on assessing and ensuring that POS attendants do not degenerate into likely offenders due to the businesses processes in use and the negligence of the POS kiosk owners. The study recommends that a large scale study is needed to ensure that further scientific findings are derived on the issues raised in the study.

Geebren, A., Jabbar, A., & Luo, M. (2021). Examining the role of consumer satisfaction within mobile eco-systems: Evidence from mobile banking services. *Computers in Human Behavior*, *114*, 106584.

Gerring, J. (2008). Case selection for case-study analysis: Qualitative and quantitative techniques. In *The Oxford handbook of political methodology*.

Gregor, S. (2006). The nature of theory in information systems. *MIS Quarterly*, 611-642.

Halford, E., Dixon, A., Farrell, G., Malleson, N., & Tilley, N. (2020). Crime and coronavirus: Social distancing, lockdown, and the mobility elasticity of crime. *Crime science*, *9*(1), 1-12.

Hasham, S., Joshi, S., & Mikkelsen, D. (2019). Financial crime and fraud in the age of cybersecurity. *McKinsey & Company*, 1-11.

Hassan, A., & Utulu, S. (2022). Socio-economic and technical factors influencing financial inclusion among indigenous peoples in Bauchi State, Nigeria. Accepted paper to be delivered at the ICTD 2022 Conference at the University of Washington in Seattle, USA, June 27-29.

Herzberg, A. (2003). Payments and banking with mobile personal devices. *Communications of the ACM*, *46*(5), 53-58.

Hindelang, M. J., Gottfredson, M. R., & Garofalo, J. (1978). *Victims of personal crime: An empirical foundation for a theory of personal victimization*. Cambridge, MA: Ballinger.

Hollis, M. E., Felson, M., & Welsh, B. C. (2013). The capable guardian in routine activities theory: A theoretical and conceptual reappraisal. *Crime Prevention and Community Safety*, *15*(1), 65-79.

Igoni, S., Onwumere, J. U. J., & Ogiri, I. H. (2020). The Nigerian digital finance environment and its economic growth: Pain or gain. *Asian Journal of Economics, Finance and Management*, 1-10.

Inegbedion, H. E. (2018). Factors that influence customers' attitude toward electronic banking in Nigeria. *Journal of Internet Commerce*, *17*(4), 325-338.

Johnson-Laird, P. N. (1999). Deductive reasoning. *Annual review of psychology*, *50*(1), 109-135.

Kasim, M., Mohammed, B., Isa, A. & Utulu, S. (2022). Developing a Social Ethical Hacking Framework for Detecting Human Factors Induced Cybersecurity Vulnerability. Paper presented at the *United Kingdom Association of Information Systems* (UKAIS 2022).

Kaur, P., Krishan, K., Sharma, S. K., & Kanchan, T. (2019). ATM card cloning and ethical considerations. *Science and engineering ethics*, *25*(5), 1311-1320.
49

Appendix 1: Interview Questions

The interview method adopted is the semi-structured method. This allowed for creativity and the emergence nature of the interview sections. It made the interview section discursive with both researcher and research participants providing through discussions issues in an emergent manner.

1. How long have you been doing your POS business?

2. How many persons have you hired

3. How do you know the right person to hire?

4. Why do your staff leave?

5. Have you ever sacked any staff? And why?

6. Are you aware of the risks attached to the POS kiosks business

7. Do you think your staff pose a challenge?

8. The processes used to complete transactions, do you think there a risks associated with it?

9. Have you ever thought about these risks before now?

10. In all how do you ensure your security and that of your customers?

Appendix 2: Examples of Empirical Evidence Supporting the Key Empirical Observation

| Observation | Construct | Comments from Interviews |
|---|---|---|
| How POS kiosk owners serve as capable guardians | Capable Guardian | *"I never thought about that and it looks like it is important to think about such things…all I think about is how to avoid customers to cheat us either with fake alert or use of magic to take our money."*<br><br>*"It is troubling that I cannot think about other ways ATM cards can be produced to reduce the open information it carries. It just too easy to steal information on ATM cards."*<br><br>*"I think that you are right. The whole problem just surfaced in my head now. If one is not careful your POS boy can actually take someone information and use it whenever he wants to."* |
| How POS kiosk attendant may evolve into likely offenders | Likely Offender | *"If one is not careful your POS boy can actually take someone information and use it whenever he wants to."*<br><br>*"I had to go the street where my boy, the one here, lives and ask about him before I hired him. The other shop, the boy there lives in my street and I know him very well. So* |

|  |  | *I didn't have any doubts about his character and integrity."* |
| --- | --- | --- |
| How POS kiosk customers may become suitable targets | Suitable Targets | *"It just too easy to steal information on ATM cards. So owners must be careful and make sure they do not fall victims."*<br><br>*"This means that customers too must request that some protocols to ensure that their cards are used in on the desks so that they can see what's going on."* |